\documentclass{article}

\usepackage[preprint,nonatbib]{nips_2018}

\usepackage[utf8]{inputenc} 
\usepackage[T1]{fontenc}    
\usepackage{hyperref}       
\usepackage{url}            
\usepackage{booktabs}       
\usepackage{amsmath,amssymb,amsfonts}       
\usepackage{nicefrac}       
\usepackage{microtype}      
\usepackage{graphicx}
\usepackage{subfigure}
\usepackage{algorithm}
\usepackage[]{algorithmic}
\usepackage{float}
\usepackage{caption}
\usepackage{multirow}
\usepackage[font=small,labelfont=bf]{caption}

\title{Quantum Walk Neural Networks for Graph-Structured Data}

\author{
  Stefan Dernbach\\
  Univeristy of Massachusetts\\
  Amherst, MA 01003 \\
  \texttt{dernbach@cs.umass.edu} \\
  \And
    Arman Mohseni Kabir\\
  Univeristy of Massachusetts\\
  Amherst, MA 01003 \\
  \texttt{arman@physics.umass.edu} \\
    \And
    Siddharth Pal\\
  Raytheon BBN Technologies\\
  Cambridge, MA 02138 \\
  \texttt{siddharth.pal@raytehon.com} \\
    \And
    Don Towsley\\
  Univeristy of Massachusetts\\
  Amherst, MA 01003 \\
  \texttt{towsley@cs.umass.edu} \\
    \And
    Miles Gepner\\
  Univeristy of Massachusetts\\
  Amherst, MA 01003 \\
  \texttt{mgepner@cs.umass.edu} \\
}

\begin{document}

\maketitle

\begin{abstract}
In recent years, new neural network architectures designed to operate on graph-structured data have pushed the state-of-the-art in the field. A large set of these architectures utilize a form of classical random walks to diffuse information. We propose quantum walk neural networks (QWNN), a novel graph neural network architecture based on quantum random walks, the quantum parallel to classical random walks. A QWNN learns a quantum walk on a graph to construct a diffusion operator which can then be applied to graph-structured data. We demonstrate the use of QWNNs on a variety of prediction tasks on graphs involving temperature, biological, and molecular datasets.
\end{abstract}

\section{Introduction}
\label{sec:Introduction}

While classical neural network approaches for structured data have been well investigated, there has been growing interest in extending neural network architectures beyond grid structured data~\cite{krizhevsky2012imagenet} 
to the domain of graph-structured data~\cite{gori2005new,scarselli2009graph,spectral,DCNN,defferrard2016convolutional,kipf2016semi}. 
In this work, we propose a quantum walk based neural network structure which can be applied to graph data.
One of the primary motivations of this work is to explore the possibility of applying quantum techniques for processing graph-structured data, and the potential advantage that it might offer over classical algorithms.
The behavior of a discrete quantum walk differs from that of a classical random walk in that the quantum walk is governed through additional operators which can be tuned to control the diffusion process in accordance to the learning task at hand. 

In addition to possible benefits of quantum inspired algorithms in tackling classical problems on classical computers, one of the major potential implications of quantum inspired machine learning algorithms is their implementation on a quantum computer. 
This paper represents a first step at developing a quantum neural network technique that can operate on graph-structured data. 
Although quantum walks have been realized using different physical systems~\cite{wang2013physical} ranging from optical lattices to molecules~\cite{du2003experimental}, the focus of this paper is the use of simulated quantum walks in a classical setting.
We show that our quantum walk based neural network approach obtains better or competitive results when compared to other state-of-the-art graph neural network approaches on a wide range of graph datasets, suggesting that quantum techniques should be further investigated in the domain of graph-structured data. 

The rest of the paper is organized as follows. Section~\ref{sec:Related Work} describes the background literature on graph neural network techniques in further detail, along with a brief literature survey on quantum walks. The setting of quantum walks on graphs is studied in Section~\ref{sec:QuantumWalks}, followed by a formal description of the proposed quantum walk neural network technique in Section~\ref{sec:QWNN}. Experimental results on node and graph regression, and graph classification tasks are described in Section~\ref{sec:Experiments}, followed by a brief discussion on the limitations of our approach in Section~\ref{sec:Limitations}. Finally, concluding remarks and a discussion on future work is presented in Section~\ref{sec:conclusion}.

\section{Related Work}
\label{sec:Related Work}

Gupta and Zia~\cite{gupta2001quantum} and Altaivsky~\cite{altaisky2001quantum} among other researchers proposed quantum versions of artificial neural networks; 
See Biamonte et al. and Dunjko et al. ~\cite{biamonte2017quantum,dunjko2017machine} for an overview of the emerging field of quantum machine learning.  While not much work exists on quantum machine learning techniques for graph structured data; in recent years, new neural networks that operate on graph structured data have sprung into prominence.
Gori et al.~\cite{gori2005new} followed by Scarselli et al.~\cite{scarselli2009graph}, proposed recursive neural network architectures to deal with graph-structured data, instead of the prevalent approach of transforming the graph data into a domain that could be handled by conventional machine learning algorithms. Bruna et al.~\cite{spectral} studied the generalization of convolutional neural networks to graph-structured signals through two approaches, one based upon a hierarchical clustering of the domain, and another based on the spectrum of the graph Laplacian. Defferrard et al.~\cite{defferrard2016convolutional} presented a CNN formulation using spectral graph theoretic methods to design fast localized convolutional filters. The method uses $k ^{th}$ order polynomials to act on $k-$hop neighborhoods of the graph, thus yielding a spatial interpretation.

Along with the spectral approaches described above, a number of spatial approaches were proposed that relied on random walks to extract and learn information from the graph. We detail two of them here for comparison. 
The first approach by Atwood and Towsley~\cite{DCNN}, is a spatial convolutional method that performs random walks on the graph and combines information from spatially close neighbors. Their approach, Diffusion Convolution Neural Networks (DCNN) use powers of the transition matrix $\mathbf{P}=\mathbf{D}^{-1}\mathbf{A}$ to diffuse information across the graph, where $A$ is the adjacency matrix, and $D$ is the diagonal degree matrix such that $\mathbf{D}_{ii}=\sum_{j}\mathbf{A}_{ij}$. The $k^{th}$ power of the transition matrix, $\mathbf{P}^k$, diffuses information from each node to every node exactly $k$ hops away from it. The output $\mathbf{Y}$ of the DCNN is a weighted combination of the diffused features from across the graph, given by
\[
\mathbf{Y}=h(\mathbf{W}\odot \mathbf{P}^*\mathbf{X}),
\]
where $\mathbf{P}^*$ is the stacked tensor of powers of transition matrices, the operator $\odot$ represents element-wise multiplication, $\mathbf{W}$ are the learned weights of the diffusion-convolutional layer, and $h$ is an activation function (i.e., reLU).

The second approach of interest due to Kipf and Welling~\cite{kipf2016semi}, was proposed to tackle semi-supervised learning using a CNN architecture built via a localized approximation of spectral graph convolutions. 
The proposed Graph Convolutional Neural Network (GCNN) works in a similar fashion as DCNN described above. Here the augmented adjacency matrix $\tilde{\mathbf{A}}=\mathbf{A}+\mathbf{I}$ and degree matrix $\tilde{\mathbf{D}}_{ii}=\sum_j \tilde{\mathbf{A}}_{ij}$ replace $\mathbf{A}$ and $\mathbf{D}$. $\tilde{\mathbf{A}}$ can be equivalently thought of as the adjacency matrix of $G$ with a self-loop on every node. The GCNN uses a doubly normalized version of $\tilde{\mathbf{A}}$ to diffuse the input with respect to the local neighborhood as follows,
\[
\mathbf{Y}=h(\tilde{\mathbf{D}}^{-1/2}\tilde{\mathbf{A}}\tilde{\mathbf{D}}^{-1/2}\mathbf{XW}),
\]
where, again, $\mathbf{W}$ are learning weights and $h$ is an activation function.

The proposed quantum walk neural network (QWNN) technique in this work is a new graph neural network architecture based on discrete quantum walks. Various researchers have worked on quantum walks on graphs -- Ambainis et al.~\cite{ambainis2001one} studied the quantum variants of random walks on one-dimensional lattices; Farhi and Gutmann~\cite{farhi1998quantum} reformulated interesting computational problems in terms of decision trees, and devised quantum walk algorithms that could solve problem instances in polynomial time compared to classical random walk algorithms that require exponential time. Aharonov et al.~\cite{aharonov2001quantum} generalized quantum walks to arbitrary graphs. Subsequently, Rohde et al.~\cite{rohde2011multi} studied the generalization of discrete quantum walks to the case of an arbitrary number of walkers acting on arbitrary graph structures, and their physical implementation in the context of linear optics. Our work uses this setting of multiple non-interacting quantum walks acting on arbitrary graphs to learn patterns in graph data.

\section{Graph Quantum Walks}
\label{sec:QuantumWalks}
\begin{figure*}
\centering
\includegraphics[scale=0.4]{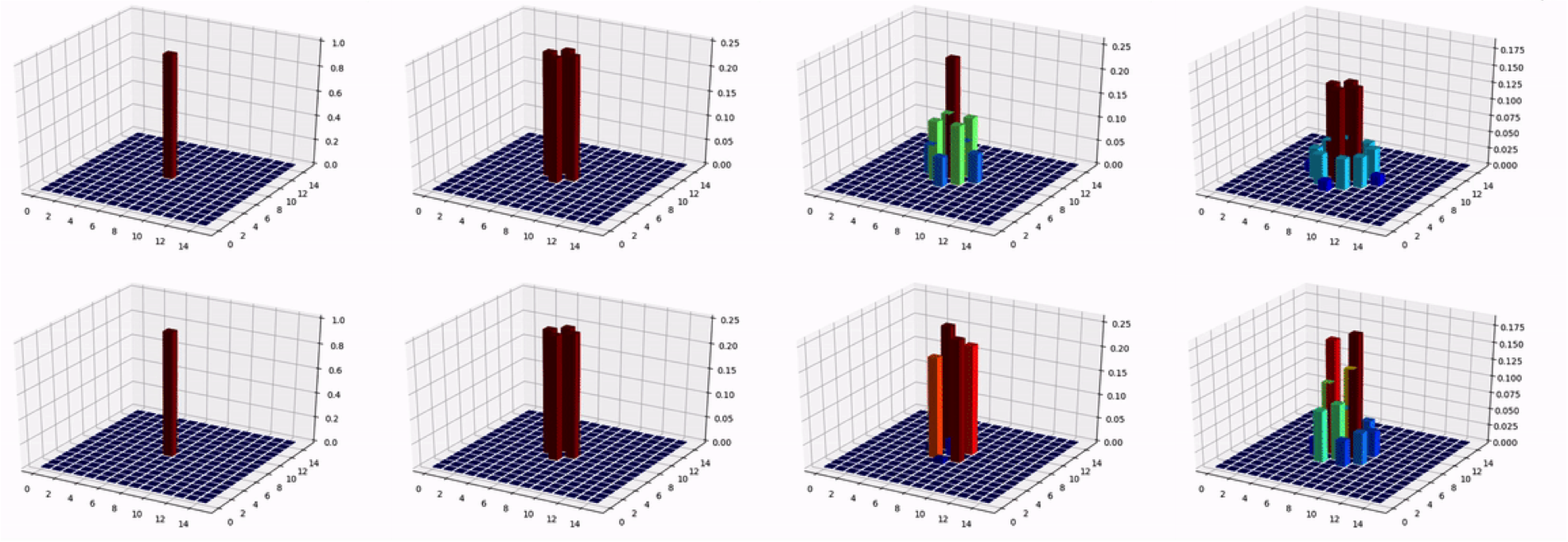}
\caption{An example demonstrating the evolution of the distribution of a classical random walk (Top Row) in comparison to a quantum random walk (Bottom Row) on a lattice graph over 4 times steps (Left to Right).}
\label{fig:quantum_evolution}
\end{figure*}

Quantum walks are a universal model for quantum computation ~\cite{childs2009universal}. They offer an alternative approach for implementing a variety of quantum algorithms, including data-base search ~\cite{shenvi2003quantum}, graph isomorphism ~\cite{douglas2008classical,qiang2012enhanced}, network analysis and navigation ~\cite{berry2010quantum,sanchez2012quantum} and quantum simulation ~\cite{schreiber20122d}. Quantum walks are the quantum parallel to their classical counterparts -- While a classical walker is modeled by a probability distribution over positions in a graph, a quantum walker is described by a superposition over position states. We utilize the form of a discrete time quantum walk on a general graph as outlined in \cite{quantumwalk,kendon2006quantum}. Given an undirected graph $G=(V,E)$, we define a position Hilbert space $\mathcal{H_P}$ spanned by basis vectors $ \left\{ \hat{\mathbf{e}} ^{(p)} _{v}, \ v \in V \right \} $ and define $\mathcal{H_{C}}$, the coin space, as an auxiliary Hilbert space of dimension $d_{max}$ spanned by the basis vectors 
$ \left\{ \hat{\mathbf{e}} ^{(c)} _{i}, \ i \in 1,\ldots,d_{max} \right \} $, where $i$ corresponds to edges incident on a vertex and $d_{max}$ is the maximum degree of the graph. For purpose of brevity, we will use $d$ instead of $d_{max}$. These states form the spin directions of a walker at vertex $v$.

A quantum walk has an associated
Hilbert space $ \mathcal{H_{W}} = \mathcal{H_{P}} \bigotimes \mathcal{H_{C}}$, which is the tensor product of position and coin Hilbert spaces. The basis states in this Hilbert space are of the form $ \hat{\mathbf{e}} ^{(p)} _{v} \otimes \hat{\mathbf{e}} ^{(c)} _{i} $, where $v \in V$ spans over set of all vertices, and $i$ is the $i^{th}$ edge incident on node $v$. 
Therefore, the position vector of a quantum walker can be written as
a linear combination of position state basis vectors,
$ \sum _{v \in V} \alpha _{v} \hat{\mathbf{e}} ^{(p)} _{v} $, for some coefficients $\{ \alpha _{v}, \ v\in V \}$ satisfying $ \sum _{v} |\alpha _v|^2 = 1 $; while the spin state at each vertex $v \in V$ can also be expressed as linear combinations of spin state basis vectors, $ \sum _{i \in 1,\ldots, d} \beta _{v,i} \hat{\mathbf{e}}  ^{(c)} _{i} $, for some coefficients $\{ \beta _{v,i}, \ i \in 1,\ldots,d \}$ satisfying $\sum _{i} \left| \beta _{v,i} \right| ^2 = 1$. The coefficients $| \alpha _v |^2$ and $| \beta _{v,i} |^2$ indicates the probability of finding the walker at vertex $v$ and the walker having spin $i$ at that vertex respectively when a measurement is done. 

A single step in the quantum random walk consists of first applying a coin operator $\mathcal{C}$ that transforms the spin state at a vertex through a unitary transformation acting on the coin Hilbert space $\mathcal{H_C}$. 
After applying the coin operator, a unitary shift operator $\mathcal{S}$ swaps the states of two vertices connected by an edge, i.e., for an edge $(u,v)$ if $u$ is the $i^{th}$ neighbor of $v$ and $v$ is the $j^{th}$ neighbor of $u$, 
then the coefficient corresponding to the basis state 
$ \hat{\mathbf{e}} ^{(p)} _{v} \otimes \hat{\mathbf{e}} ^{(c)} _{i} $ gets swapped with that for the basis state $ \hat{\mathbf{e}} ^{(p)} _{u} \otimes \hat{\mathbf{e}}  ^{(c)} _{j} $. 
Let $\Phi ^{(t)}$ in $\mathcal{H_W}$ denote the superposition state of the quantum walker at time $t$.
If $\Phi ^{(0)}$ is the initial state of the quantum walker on $G$ then after $t$ time steps the state of the walker is described by: $\Phi ^{(t)}$=$U^{t} \Phi ^{(0)}$, where a single step of the discrete quantum walk on graph $G$ can be expressed in shorthand as $U=\mathcal{S}(\mathcal{C}\otimes I_{|V|})$. 
Note that the quantum walks  behave very differently from classical random walks in that they can be heavily influenced by the choice of the initial superposition as well as the coin operator. This allows more degrees of freedom to the deep learning technique to fit the data through a controlled diffusion process.
Figure~\ref{fig:quantum_evolution} shows how the quantum walk differs from a classical walk over a lattice in that the probability distribution of the classical walk is uniform over space which is not true for the quantum walk.

\section{Quantum Walk Neural Networks}
\label{sec:QWNN}

\begin{figure*}
\centering
\includegraphics[scale=0.6]{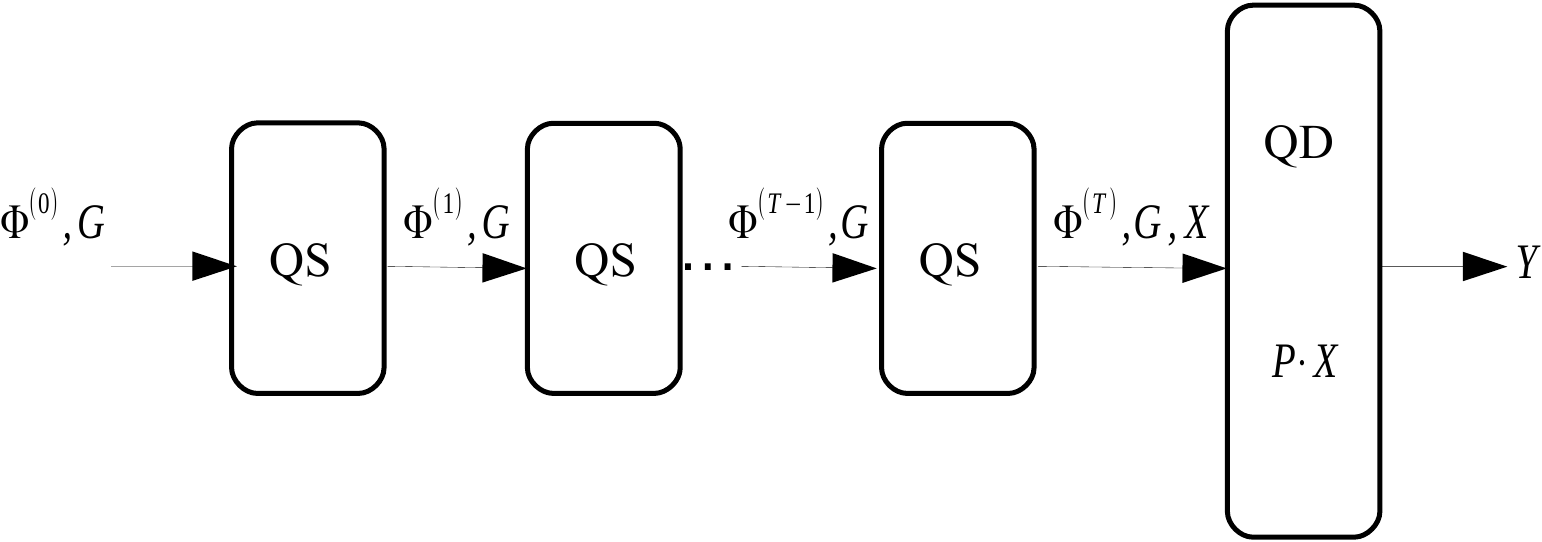}
\caption{A schematic diagram of the QWNN architecture. The initial superposition $\Phi ^{(0)}$, and graph $G$, are fed to the quantum step (QS) layers to obtain the final superposition $\Phi ^{(T)}$, which in turn is fed to the quantum diffusion (QD) layer along with the feature matrix $X$ to obtain the output $Y$.  }
\end{figure*}

Our proposed neural network architecture, Quantum Walk Neural Networks (QWNN), learns a quantum random walk on a graph by means of learning the coin operators and/or the initial superposition of the walker. We further suggest two separate implementations of a QWNN based on whether the coin operators are defined spatially or temporally. In the spatial setting there is a separate coin operator for each node in the graph. In the temporal case, every node in the graph shares the same coin operator, but that operator is unique to each step in the walk. The first setting allows the quantum walk to behave differently in different sections of the graph while the latter setting allows the walk dynamics to evolve over time.

A QWNN then uses the walker distribution induced by the quantum random walk to form a diffusion operator which act on an input feature matrix. Given a tensor $\mathbf{\Phi}^{(0)}\in \mathbb{R}^{W \times N\times d}$ representing a set of $W$ walkers and their corresponding initial superpositions, a set of spatial coins $\mathbf{C}\in \mathbb{R}^{N \times d \times d}$ or temporal coins $\mathbf{C}\in \mathbb{R}^{T \times d \times d}$ and a shift tensor $\mathbf{S}\in \mathbb{Z} _2 ^{N \times d \times N \times d}$, a quantum walk neural network takes as input a matrix of $F$ features per node $\mathbf{X}\in\mathbb{R}^{N\times F}$, and outputs diffused features $\mathbf{Y}\in \mathbb{R}^{N\times F}$. Note that, when spatially distributing coins, the dimension of the coin on each node equals its degree, but for sake of implementation we use $d$ universally and set the extra coin variables as zero. For temporally distributed coins, the graph is made $d$-regular by adding self loops to nodes.

The network architecture is composed of a sequence of quantum step layers leading to a diffusion layer. Each quantum step layer takes as input the shift tensor which encodes the graph structure and the current superposition tensor. The layer evolves the superposition as follows: For an edge $(u,v)$, with $u$ being the $i ^{th}$ neighbor of $v$ and $v$ being the $j^{th}$ neighbor of $u$, $\mathbf{\Phi}^{(t+1)} _{wuj}= \left( \mathbf{\Phi}^{(t)} \mathbf{C} \right) _{wvi}$ and $\mathbf{\Phi}^{(t+1)} _{wvi}= \left( \mathbf{\Phi}^{(t)} \mathbf{C} \right) _{wuj}$. Hence, we set the shift tensor $\mathbf{S}$ as $S_{ujvi}=1$ iff $u$ is the the $i ^{th}$ neighbor of $v$ and $v$ is the $j^{th}$ neighbor of $u$. The output $\mathbf{\Phi}^{(t+1)}$ is fed into either the next quantum step layer or a final diffusion layer. The diffusion layer accepts a superposition tensor $\mathbf{\Phi}$ transforms it into a diffusion matrix $\mathbf{P}$ by summing the squares of the spin states at each vertex to calculate a transition probability. It then uses $\mathbf{P}$ and the input feature matrix $\mathbf{X}$ and outputs a new feature matrix $\mathbf{Y}$.

For the first layer in the network, we initialize $\mathbf{\Phi}^{(0)}$ by constructing a unique walker at each node in the graph with equal spin along each incident edge. If we are learning the initial amplitudes, $\mathbf{\Phi}^{(0)}$ is stored as a variable that is updated during backpropagation. The method for a forward pass of the complete network (using spatial coins) is given in Algorithm \ref{alg:qwnn}. The equation $\mathbf{\Phi}^{(t)}_{\cdot i\cdot}\gets \mathbf{\Phi}^{(t-1)}_{\cdot i\cdot} \cdot \mathbf{C}_{i \cdot \cdot}$ is replaced with $\mathbf{\Phi}^{(t)}_{\cdot i\cdot}\gets \mathbf{\Phi}^{(t-1)}_{\cdot i\cdot} \cdot \mathbf{C}_{t \cdot \cdot}$ in the temporal coin case. The notation $\mathbf{A\cdot B}$ denotes the inner product between tensors $\mathbf{A}$ and $\mathbf{B}$, the operator $\mathbf{A:B}$ is the inner product of $\mathbf{A}$ and $\mathbf{B}$ over two dimensions, and $\mathbf{A\odot B}$ is an elementwise product. The final diffusion equation $\mathbf{Y} \gets \mathbf{P}\mathbf{X}$ is similar to other classical graph neural networks that use a random walk diffusion operation. However there is no explicit weight matrix in the network because it is implicit in learning the random walks.

\begin{algorithm}
	\caption{QWNN Forward Pass}
	\label{alg:qwnn}
	\begin{algorithmic}
		\STATE \textbf{given} Initial Superpositions $\mathbf{\Phi}^{(0)}$, Coins $\mathbf{C}$, Shift $\mathbf{S}$
		\STATE \textbf{input} Features $\mathbf{X}$
		\FOR{$t=1$ to $T$}
		\FOR{All nodes $v_i$}
		\STATE $\mathbf{\Phi}^{(t)}_{\cdot i\cdot}\gets \mathbf{\Phi}^{(t-1)}_{\cdot i\cdot} \cdot \mathbf{C}_{i \cdot \cdot}$\\
		\ENDFOR
		\STATE 
        $\mathbf{\Phi}^{(t)} \gets \mathbf{\Phi}^{(t)}: \mathbf{S}$ \ \ 
           \big(i.e., $ \mathbf{\Phi} ^{(T)} _{wuj} = \sum _{v} \sum _{i}   
          \mathbf{\Phi} ^{(T)} _{wvi} S_{viuj} $ \big)  \\
		\ENDFOR
		\STATE $\mathbf{P} \gets \sum_\text{k} \mathbf{\Phi}^{(T)}_{\cdot \cdot k}\odot\mathbf{\Phi}^{(T)}_{\cdot \cdot k}$ \\
		\STATE $\mathbf{Y} \gets h(\mathbf{P}\mathbf{X}+\mathbf{b})$\\
		\textbf{return} $\mathbf{Y}$
	\end{algorithmic}
\end{algorithm}

Previous work on quantum walks \cite{quantumwalk} use Grover's diffusion operator as the coin operator because it is the only nontrivial, real-valued transform that is (1) unitary and (2) treats all edges connected to a vertex identically. The first requirement guarantees that superposition of each walker retains unit norm. The second restriction is added in order to prevent edge ordering from affecting the walk. In a QWNN we can choose to relax these conditions for certain tasks to allow for learning biased coin operators. Removing the unitary constraint allows us to learn expansive or contractive diffusion operators. In the spatial coin setting the learning can adapt to any edge ordering equally so the second condition is unnecessary. 

\subsection{Unitary Operators}
When the unitary property of the coin operator is desirable, we can initialize each coin to any unitary matrix. However in general, gradient updates to the coin operators cause them to leave the space of unitary matrices. The result of this is that the matrix $\mathbf{P}$ is no longer a probability matrix and $P_{uv}$ no longer describes the likelihood of a walker beginning on node $u$ and ending on node $v$ of the graph.

Unitary constraints are common in deep recurrent neural networks
because nonunitary weight matrices can lead to vanishing or exploding gradients.
In our implementation, we use the general parametrization method in \cite{jing2016tunable} which can represent a large subset of unitary matrices.
A $N \times N$ unitary matrix $\mathbf{U}_{N}$ is represented as a product of rotation matrices $\mathbf{R}_{k\ell}$ and a unitary diagonal matrix $\mathbf{D}$, such that 
$\mathbf{U}_{N} = \mathbf{D}\prod_{k=2}^{k=N}\prod_{\ell=1}^{\ell=i-1}\mathbf{R}_{k\ell}$ where $\mathbf{R}_{k\ell}$ is defined as the $N$-dimensional identity matrix with the elements $R_{kk}$, $R_{k\ell}$ , $R_{\ell k}$ and $R_{\ell \ell}$ replaced as follows:

\[\begin{pmatrix}
R_{kk} & R_{k \ell} \\  
R_{\ell k} & R_{\ell \ell}
\end{pmatrix}= \begin{pmatrix}
e^{i\varphi_{k \ell}}\cos(\theta_{k \ell}) & -e^{i\varphi_{k\ell}}\sin(\theta_{k\ell}) \\  
\sin(\theta_{k\ell}) & \cos(\theta_{k\ell})
\end{pmatrix}\]

with $\theta_{k \ell}$ and $\varphi_{k \ell}$ being parameters specific to $\mathbf{R}_{k \ell}$ which can be updated through backpropagation.
Each of these rotation matrices acts as an unitary transformation on a 2-dimensional subspace of the $N$-dimensional Hilbert space and leaves the remaining $N-2$ dimensions unchanged. 

\subsection{Edge Ordering}
\label{sec:edges}
When sharing coin operators across nodes, edge ordering becomes a factor in determining the span of possible quantum walks. This is easily demonstrated in a simple 3-node cycle graph. If each node has a defined right and left neighbor, such that always following the right or left neighbor will allow one to tour the graph, then we can easily define the coin operator
$
\mathbf{C}=\begin{pmatrix}
0 & 1\\
1 & 0
\end{pmatrix}
$
that will cause a quantum walker to follow this tour. However, if one node has the order of its two edges reversed, no such quantum walk is definable using a single coin.

To address this issue we propose two heuristic approaches to edge ordering. The first is a global ordering of nodes based on their betweeness centrality. The betweeness centraility~\cite{brandes2001faster}
of node $v_i$ is calculated as:
\[
g(v_i)=\sum_{j\neq i \neq k}\frac{\sigma_{jk}(v_i)}{\sigma_{jk}}
\]
where $\sigma_{jk}$ is the number of shortest paths from $v_j$ to $v_k$ and $\sigma_{jk}(v_i)$ is the number of shortest paths from $v_j$ to $v_k$ that pass through $v_i$. This ranks all nodes and an edge connected to a higher ranked node will always precede an edge connected to a lower ranked node. In this setting, a walker moving along a higher ranked edge is moving towards a more central part of the graph compared to a walker moving along a lower ranked edge.

The second approach is a local edge ordering where a similarity score between every pair of adjacent nodes is computed and the edges incident to each node are ordered according to the score of the two nodes it connects. This ordering does not induce a global ranking of nodes, each node has its own ranking of neighbors.
This approach allows the quantum walk to distinguish between transitioning to more or less similar nodes in the graph. We use a random walk node similarity measure defined as 
\[
S(v_i,v_j)=\mathbf{W}^k_i \left(\mathbf{W}^k_j \right)^T
\]
where $\mathbf{W}^k$ is a (classical) random walk matrix raised to the $k$th power. 

Neither of these methods guarantee that there will not be a need to break ties between edges. However, in practice, these ties often come from symmetries in the graph and do not have a real affect on the global behavior of the quantum walk.

\section{Experiments}
\label{sec:Experiments}
We demonstrate the flexibility of our approach across a wide range of tasks. Quantum Walk Neural Networks prove an effective method for regression tasks at both the node and graph level as well as graph classification.

\subsection{Node Regression Task}


\begin{figure}
\centering	\includegraphics[width=0.7\linewidth,height=0.20\textheight]{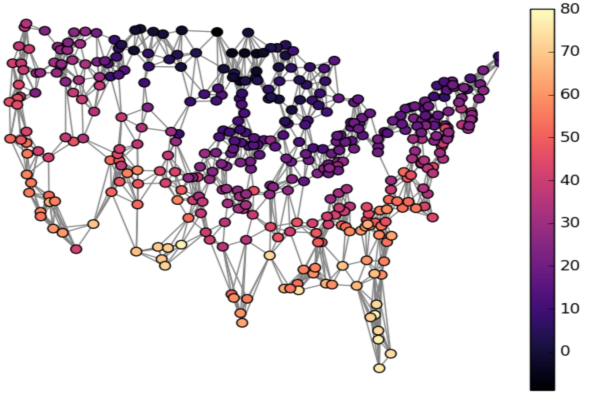}
	\caption{An example day's temperature readings. The depicted graph is constructed from the 8 nearest geographical neighbors of each node.}
	\label{fig:ustemp}
\end{figure}


We demonstrate the use of our network in learning to predict daily temperatures. The data consists of a years worth of daily high temperatures recorded at 409 locations across the United Sates in 2009 \cite{climate}.
A nearest neighbors graph from the stations' longitudes and latitudes using 8 neighbors is constructed, which was empirically found to produce a fully connected graph  (Fig.\ref{fig:ustemp}). The high temperature from each station on a single day is then used to predict the following day's high temperatures. We form our neural network from a series of quantum step layers (indicated by walk length) and a single diffusion layer. The QWNN uses the spatial coin setup and due to the sum of norm of the temperature vector changing from day to day, we do not restrict the coins to be unitary for this experiment.
We compare against a diffusion convolution neural network (DCNN) \cite{DCNN} of varying walk lengths and the graph convolution neural net (GCNN) in \cite{kipf2016semi}. The GCNN does not take as input a walk length or number of hops and the size of the learnable weight matrix $F_{in}\times F_{out}$ is not well suited when the in and out dimensions are both only $1$. We show results for it regardless for completeness. 
The data is divided into thirds for training, validation, and testing and the learning is limited to 128 epochs or early stopping if the validation score stops improving. 

Table. \ref{tab:res} gives the test results for the trained networks. Both the mean and standard deviation are reported taken from 5 trial runs per network configuration. We observe that the quantum walk techniques yield superior regression results compared to the GCNN and DCNN techniques.
We also note that learning both the coin operators and the initial amplitudes gives a minimal improvement over learning only one of the two. Additionally in most cases, learning the coin operators while keeping the initial amplitudes constant is more effective than learning the initial amplitudes but keeping the coin operators constant. Surprisingly, learning the amplitudes shows a marginal improvement over learning the coin despite the coin operators having a larger potential effect on the walk. One possible explanation is that the reduced number of parameters for the amplitudes could make learning easier.
For quantum random walk networks, longer walks show an improvement over shorter walks while the DCNNs do not exhibit the same trend. 
\begin{table}
\centering
\caption{Temperature Prediction Results}
\label{tab:res}
  \begin{tabular}{|c|r|r|r|}
    \hline
    \multicolumn{4}{|c|}{RMSE $\pm$ STD} \\ \hline
    \multicolumn{1}{|l|}{\multirow{2}{*}{}} & \multicolumn{3}{c|}{Walk Length}\\ \cline{2-4} 
    \multicolumn{1}{|l|}{} & $2$ 				 & $3$ 				 & $4$               \\ \hline
    GCNN                & $8.56\pm0.02$    &                   &               \\ \hline
    DCNN 						& $7.40\pm 0.13$   & $7.46\pm 0.06$ & $7.44\pm 0.10$ \\ \hline
    QW (A+C)  & $5.54\pm 0.16$  & $5.38 \pm 0.07$ & $5.28\pm 0.08$  \\ \hline
    QW (A)                  & $5.43\pm 0.15$ & $5.43\pm 0.04$   & $5.42\pm 0.23$ \\ \hline
    QW (C)                 & $6.05\pm 0.27$  & $5.85\pm 0.07$  & $5.97\pm 0.12$  \\ \hline
  \end{tabular}
\end{table}

\subsection{Graph Classification Tasks}
We apply QWNNs to several common graph classification datasets: Enzymes \cite{enzymes}, Mutag \cite{mutag}, and NCI1 \cite{nci1}. Enzymes is a set of 600 molecules extracted from the Brenda database \cite{brenda}. The task is to categorize each enzyme into 1 of 6 classes. Mutag is a dataset of 188 mutagenic aromatic and heteroaromatic nitro compounds that are classified into 1 of 2 categories, mutagenic or nonmutagenic. NCI1 consists of 4110 graphs representing two balanced subsets of datasets of chemical compounds screened for activity against nonsmall cell lung cancer. Summary statistics for each dataset are given in Table \ref{tab:dsets}.

For the QWNN and GCNN, we construct a neural network with the graph layer, a hidden dense layer of size $10$ and an output layer with a softmax activation. The results from DCNN are taken from \cite{DCNN}. The QWNN uses a walk of length 4 and a single coin per timestep. Table \ref{tab:dsets} report results for QWNN using the two different edge ordering given in Section \ref{sec:edges}. QWNNs demonstrate the highest mean accuracy on 2 out of the 3 datasets though GCNNs are within the margin of error. The reverse is true on the Enzymes dataset. On this experiment, both QWNNs and GCNNs have a training accuracy of $0.7+$ but do not generalize well to the test set.

\begin{table}
\centering
\caption{Graph Classification Datasets Summary and Results}
\label{tab:dsets}
  \begin{tabular}{|c|c|c|c|}
    \hline
     				& Enzymes 	& Mutag & NCI1 \\ \hline
    Graphs 			& $600$     & $188$ & $4110$ \\ \hline
    Average Nodes	& $33$		& $18$ 	& $30$ \\ \hline
    Max Nodes		& $126$    	& $28$ 	& $111$ \\ \hline
    Max Degree 		& $9$  		& $4$ 	& $4$ \\ \hline
    Node Classes 	& $3$		& $7$	& $37$\\ \hline
    Graph Classes	& $6$     	& $2$ 	& $2$ \\ \hline
    & \multicolumn{3}{|c|}{Classication Accuracy $\pm$ STD} \\ \hline
 GCNN 			& $0.41\pm 0.08$	& $0.83\pm 0.09$	& $0.70\pm 0.02$			\\ \hline
 DCNN 			& $0.19$ 			& $0.67$ 	& $0.63$ 	\\ \hline
 QWNN (centrality) 	& $0.32\pm 0.03$	& $0.85\pm 0.06$			& $0.68\pm 0.02$\\ \hline
 QWNN (similarity)	& $0.32\pm 0.03$	& $0.92\pm 0.03$			& $0.71\pm 0.01$\\ \hline
  \end{tabular}
\end{table}



\subsection{Graph Regression Task}
The QM7 dataset\cite{blum,rupp} is a collection of 7165 molecules each containing up to 23 atoms. The geometries of these molecules are stored in Coulomb matrix format defined as 

\[\mathbf{C}_{ij}=
\left\{\begin{matrix}
&0.5Z_i^{2.4}  \quad i=j & \\ 
&\frac{Z_i Z_j}{|R_i-R_j|}  \quad i\neq j & 
\end{matrix}\right.\]

where $Z_i$, $R_i$ are the charge of and position of the $i$th atom in the molecule respectively. The goal of the task is to predict the atomization energy of each molecule. Atomization energies of the molecules range from -440 to -2200 kcal/mol.

For our task, we form an approximation of the molecular graph from the Coulomb matrix by normalizing out the charges and separating the distances between atoms into 2 sets using K-means. One set contains the atom pairs with larger distances between them and the other the smaller distances. We create an adjacency matrix from all pairs of atoms in the smaller distance set. This creates connected graphs for all but 19 of the molecules. For those remaining 19 molecules, we iteratively add more distant pairs of atoms until the molecular graph is fully connected. We take this approach to constructing the graphs because most molecules have a noticeable distance gap between directly bonded and unbonded pairs of atoms. We use the element of each as atom encoded as a one-hot vector as the input features on each node. 

Our neural network is composed of quantum random walk layer of length 4 using 1 unitary coin per time step. Graphs that are smaller than the largest molecules (23 atoms) have both their feature and adjacency matrices appended with 0s to bring them up to size. The quantum walk layer feeds into two hidden layers of size 400 and 100 respectively and a final layer that outputs a single value with a sigmoid activation which is then rescaled to the range of atomization energies.

For comparison we use the same 5-fold division of the data for training and testing as in previous work. In addition to the other graph neural network approaches, we compare against a multilayer neural network \cite{montavon2012learning}. This network takes as input 1800 binary features derived from the coulomb matrix in comparison to the single atom class we use as input.

The RMSE and MAE with a standard deviation are given in Table \ref{tab:atom}. QWNNs give noticably better results over the other graph neural network approaches however, there is obvious room to improve graph NN techniques as evidenced by the lower error of the Multilayer NN.

\begin{table}
  \centering
  \caption{Atomization Energy Prediction Results}
  \label{tab:atom}
  \begin{tabular}{|c|r|r|}
    \hline
                            & $RMSE$                & $MAE$ \\ \hline
    GCNN					& $19.26 \pm 0.47$		&	$14.77 \pm 0.26$\\ \hline
    DCNN  		& $15.51 \pm 1.12$      & $10.73 \pm 0.35$\\ \hline
    QWNN (centrality) 		& $12.52 \pm 0.91$      & $8.93 \pm 0.14$\\ \hline    
    QWNN (similarity) 		& $16.01\pm 0.96$      & $10.78 \pm 0.22$\\ \hline   
    \hline
    Multilayer NN 			& $5.96 \pm 0.48$  		& $3.51\pm 0.13$\\ \hline
  \end{tabular}
\end{table}

\section{Limitations}
\label{sec:Limitations}
Storing the superposition of a walker requires $O(Nd)$ space, with $N$ the number of nodes in the graph, and $d$ the max degree of the graph. To calculate a complete diffusion matrix requires a walker to begin at every node, and for the superposition of the walkers to be tracked for $T$ time steps. This leads to a space requirement of $O(N^2dT)$ which is intractable for very large graphs, especially when doing learning on a GPU. 

Quantum walks with shared coins across nodes are not invariant to the ordering of the edges to each node. In practice, we find that our heuristic ordering works well, Nonetheless, a QWNN can perform differently on isomorphic graphs. The spatial coin implementation of the QWNN does not inherit this problem because each coin can be tuned to the edge ordering specific to the node it operates at.

\section{Concluding Remarks}
\label{sec:conclusion}
Quantum walk neural networks offer a noticeable improvement over standard graph neural networks in both node and graph regression tasks and are competitive in classification tasks. QWNNs are able to adapt both spatially across the graph or temporally over the walk. The final diffusion matrix after learning has converged, can be cached for quick forward passes in the network. However the constant coin and shift operations during learning lead to a marked slowdown in our architecture compared to others, with space complexity also being an issue. In the current work, each walker on the graph operates independently. A future research direction is to investigate learning multi-walker quantum walks on graphs. Reducing the number of independent walkers and allowing interactions can reduce the space complexity of the quantum step layers. 

\bibliography{main}
\bibliographystyle{plain}

\end{document}